\documentclass{emulateapj}
\usepackage{graphicx,color}
\usepackage{amssymb,amsmath}
\usepackage[colorlinks,hyperfootnotes=false,citecolor=blue,linkcolor=blue]{hyperref}

\newcommand{\RB}{\textcolor{blue}{RB15}}
\newcommand{\Oesch}{\textcolor{blue}{O15}}
\newcommand{\Finkelstein}{\textcolor{blue}{F13}}

\begin{document}
\shorttitle{Spectroscopic Measurement of a Redshift z=8.68 Galaxy}
\shortauthors{Zitrin et al.}

\slugcomment{Submitted to the Astrophysical Journal Letters}

\title{Lyman-alpha Emission from a Luminous  \lowercase{z}=8.68 Galaxy: Implications for Galaxies as Tracers of Cosmic Reionization}


\author{Adi Zitrin\altaffilmark{1,2}, Ivo Labb\'e\altaffilmark{3}, Sirio Belli\altaffilmark{1}, Rychard Bouwens\altaffilmark{3}, Richard S. Ellis\altaffilmark{1}, Guido Roberts-Borsani\altaffilmark{3,4}, Daniel P. Stark\altaffilmark{5}, Pascal A. Oesch\altaffilmark{6,7}, Renske Smit\altaffilmark{8}}
\altaffiltext{1}{Cahill Center for Astronomy and Astrophysics, California Institute of Technology, MC 249-17, Pasadena, CA 91125, USA; adizitrin@gmail.com}
\altaffiltext{2}{Hubble Fellow}
\altaffiltext{3}{Leiden Observatory, Leiden University, NL-2300 RA Leiden, The Netherlands}
\altaffiltext{4}{Department of Physics and Astronomy, University College London, Gower Street, London WC1E 6BT, UK}
\altaffiltext{5}{Steward Observatory, University of Arizona, 933 N Cherry Ave, Tucson, AZ 85721 USA}
\altaffiltext{6}{Yale Center for Astronomy and Astrophysics, Physics Department,New Haven, CT 06520, USA}
 \altaffiltext{7}{Department of Astronomy, Yale University, New Haven, CT 06520, USA}
 \altaffiltext{8}{Centre for Extragalactic Astronomy, Department of Physics, Durham University, South Road, Durham DH1 3LE, UK}


\begin{abstract}
We report the discovery of Lyman-alpha emission (Ly$\alpha$) in the bright galaxy EGSY-2008532660 (hereafter EGSY8p7) using the MOSFIRE spectrograph at the Keck Observatory. First reported by \citet{Borsani2015RedIRAC}, it was selected for spectroscopic observations because of its photometric redshift ($z_{phot}=8.57^{+0.22}_{-0.43}$), apparent brightness (H$_{160}=25.26\pm0.09$) and red Spitzer/IRAC [3.6]-[4.5] color indicative of contamination by strong oxygen emission in the [4.5] band. With a total integration of $\sim$4.3 hours, our data reveal an emission line at $\simeq$11776 {\AA} which we argue is likely Ly$\alpha$ at a redshift $z_{spec}=8.683^{+0.001}_{-0.004}$, in good agreement with the photometric estimate. The line was detected independently on two nights using different slit orientations and its detection significance is $\sim7.5\sigma$. An overlapping skyline contributes significantly to the uncertainty on the total line flux although the significance of the detected line is robust to a variety of skyline-masking procedures. By direct addition and a Gaussian fit, we estimate a 95\% confidence range of 1.0--2.5$\times10^{-17}$ erg s$^{-1}$ cm$^{-2}$, corresponding to a rest-frame equivalent width of 17--42 {\AA}. EGSY8p7 is the most distant galaxy confirmed spectroscopically to date, and the third luminous source in the EGS field beyond $z_{phot}\gtrsim7.5$ with detectable Ly$\alpha$ emission viewed at a time when the intergalactic medium is believed to be fairly neutral. Although the reionization process was probably patchy, we discuss whether luminous sources with prominent IRAC color excesses may harbor harder ionizing spectra than the dominant fainter population thereby creating earlier ionized bubbles. Further spectroscopic follow-up of such bright sources promises important insight into the early formation of galaxies. \vspace{0.05cm}
\end{abstract}

\keywords{cosmology: observations --- galaxies: high-redshift --- galaxies: evolution --- galaxies: formation}

\section{Introduction}\label{intro}
Our understanding of the early Universe has improved considerably in recent years through the photometric discovery of large numbers of high-redshift galaxies in both deep and gravitationally-lensed fields observed with the \emph{Hubble Space Telescope} \citep[e.g.][]{Ellis2013Highz,McLure2013,Bradley2013highz,Oesch2014LumZ910,Zheng2014A2744,Bouwens2015LF10000}. Although uncertainties remain, the demographics and limited spectroscopic follow-up of this early population has been used to argue that star-forming galaxies played a significant role in completing the reionization of the IGM by redshift $z\sim6$ \citep[e.g.][]{Robertson2013highzhudf, Robertson2015, KuhlenFaucher2012Reinozation, Bouwens2012Reionization,Bouwens2015Planck, Finkelstein2012, Finkelstein2015, Ishigaki2015}. A key observation delineating the end of the reionization epoch is the marked decline beyond $z\simeq$6 in the visibility of Lyman-alpha (Ly$\alpha$) emission seen in continuum-selected galaxies \citep{Stark2010z3-7fractions,Fontana2010reionizationLya,Ono2012,Pentericci2011Lyfractionz7,Schenker2012DeclinLyAinZ,Schenker2014above7Spec}. As Ly$\alpha$ represents a resonant transition it is readily scattered by the presence of neutral gas and thus acts as a valuable proxy for the state of the IGM.

\begin{figure*}
 \begin{center}
   \includegraphics[width=170mm,trim=4cm 3.5cm 0cm 3cm,clip]{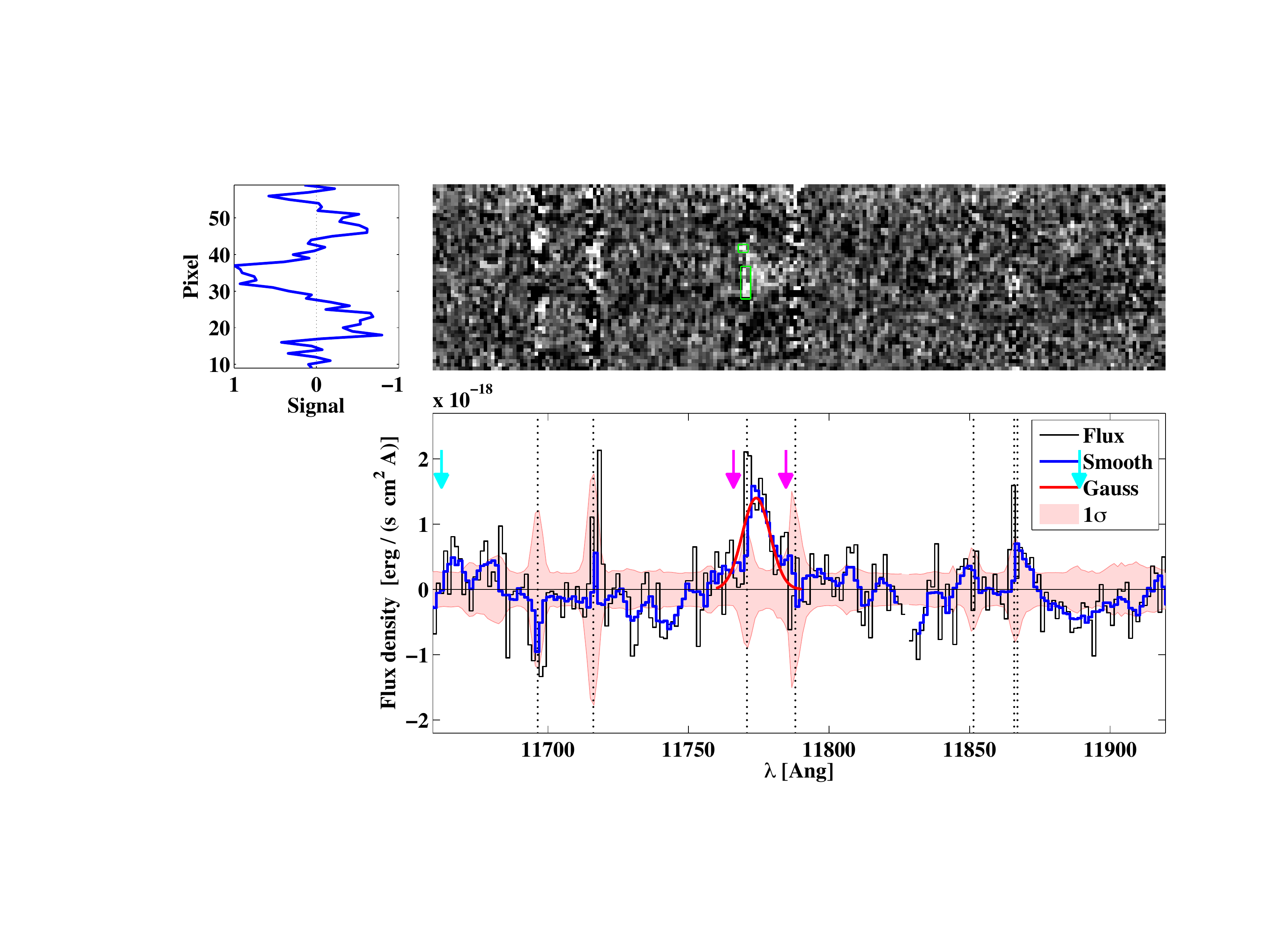}
 \end{center}
\caption{Spectroscopic detection of emission in EGSY8p7 with MOSFIRE. \emph{Upper panel} shows the 2D spectrum below which we plot the raw (\emph{black line}) and smoothed (\emph{blue line}) 1D spectrum and its error (\emph{red shading}). The \emph{red line} shows an example best-fit model of the data (\S \ref{lya}).Vertical lines mark OH skyline positions. The \emph{upperleft} panel shows a normalized signal map extracted along the slit within a 5-pixel ($\simeq6.5$ {\AA}) wide box centered on the line. The pattern of two negative peaks bracketing the positive peak exactly matches that expected from the dithering scheme used. \emph{Arrows} show the predicted locations of other lines for a lower redshift interpretation of the line. \emph{Green} boxes on the 2D spectrum mark the skyline region typically masked out in our calculations. See \S \ref{lya} for more details.}\vspace{0.1cm}
\label{fig1}
\end{figure*}

There are, however, several limitations in using Ly$\alpha$ as a probe of reionization. Firstly, converting the declining visibility of Ly$\alpha$ into a neutral gas fraction involves complex radiative transfer calculations and several uncertain assumptions. The currently observed decline \citep[e.g.][]{Schenker2014above7Spec} implies a surprisingly rapid end to the process \citep{Choudhury2014,Mesinger2015reionizationLya}. Secondly, simulations suggest that reionization is likely to be a patchy process \citep{TaylorLidz2014reionization} and thus conclusions drawn from the modest samples of spectroscopically-targeted galaxies may be misleading \citep{Treu2012z7-8,Pentericci2014z7,Tilvi2014RapidDecline}. Finally, the absence of a Ly$\alpha$ detection in the spectrum of a proposed high-redshift candidate may simply imply the source is a foreground galaxy (although the contamination rate for such galaxies is typically very low, e.g. \citealt{Vanzella2009LyAlpha,Stark2010z3-7fractions}). Ideally targets in such studies would be spectroscopically confirmed independently of Ly$\alpha$, for example using UV metal lines \citep[][see also \citealt{Zitrin2015CIII}]{Stark2015CIIIdetectionz67,Stark2015CIV}.

Although observational progress is challenged by the faintness of targets selected in deep fields such as the Hubble Ultra Deep Field (typically $m_{AB}\sim$27-28), an important development has been the identification of much brighter $z>7$ candidates from the wider area, somewhat shallower, Cosmic Assembly Near-infrared Deep Extragalactic Legacy Survey \citep[CANDELS,][]{Grogin2011Candles,Koekemoer2011Candles}. Surprisingly, some of these brighter targets reveal Ly$\alpha$ despite lying inside the putative partially neutral era. \citet[][hereafter \Finkelstein]{Finkelstein2013} reported Ly$\alpha$ with a rest-frame equivalent width (EW) of 8 \AA\ at $z$=7.508 in a $H_{AB}$=25.6 galaxy; \citet[][hereafter \Oesch]{Oesch2015z773} find Ly$\alpha$ emission at $z$=7.73 with EW=21 \AA\ in an even brighter source at $H_{AB}$=25.03; and \citet[][hereafter \RB ]{Borsani2015RedIRAC} identified a tentative Ly$\alpha$ emission (4.7$\sigma$) in a $H_{AB}$=25.12 galaxy at a redshift $z$=7.477, which we have now confirmed (Stark et al, in prep). In addition to their extreme luminosities ($M_{UV}\simeq-22$), these three sources have red [3.6]-[4.5] Spitzer/IRAC colors, indicative of contamination from strong [O III] and Balmer H$\beta$ emission. 

Using the Multi-Object Spectrometer For Infra-Red Exploration (MOSFIRE, \citealt{KeckMOSFIRERef}) on the Keck 1 telescope, we report the detection of a prominent emission line in a further bright candidate drawn from the CANDELS program. EGSY-2008532660 (hereafter EGSY8p7; RA=14:20:08.50, DEC=+52:53:26.60) is a $H_{AB}$=25.26 galaxy with a photometric redshift of $8.57^{+0.22}_{-0.43}$ and a red IRAC [3.6]-[4.5] color, recently discovered by \RB. We discuss the likelihood that the line is Ly$\alpha$ at a redshift $z_{spec}=8.68$ making this the most distant spectroscopically-confirmed galaxy. Detectable Ly$\alpha$ emission at a redshift well beyond $z\simeq8$ raises several questions regarding both the validity of earlier claims for non-detections of Ly$\alpha$ in fainter sources, and the physical nature of the luminous sources now being verified spectroscopically. Even if these bright systems are not representative of the fainter population that dominate the ionization budget, they offer new opportunities to make spectroscopic progress in understanding early galaxy formation.

The paper is organized as follows: In \S \ref{obs} we review the object selection, spectroscopic observations, and data reduction. The significance of the line detection and its interpretation as Ly$\alpha$ is discussed in \S \ref{lya}. We discuss the implications of the detectability of Ly$\alpha$ in the context of the earlier work in \S \ref{discussion}. Throughout we use a standard $\Lambda$CDM cosmology with $\Omega_{\rm m0}=0.3$, $\Omega_{\Lambda0}=0.7$, $H_{0}=100$ $h$ km s$^{-1}$Mpc$^{-1}$, $h=0.7$, and magnitudes are given using the AB convention. Errors are $1\sigma$ unless otherwise stated.

\begin{figure*}
 \begin{center}
   \includegraphics[width=175mm,trim=1.3cm 0.5cm 4cm 0cm,clip]{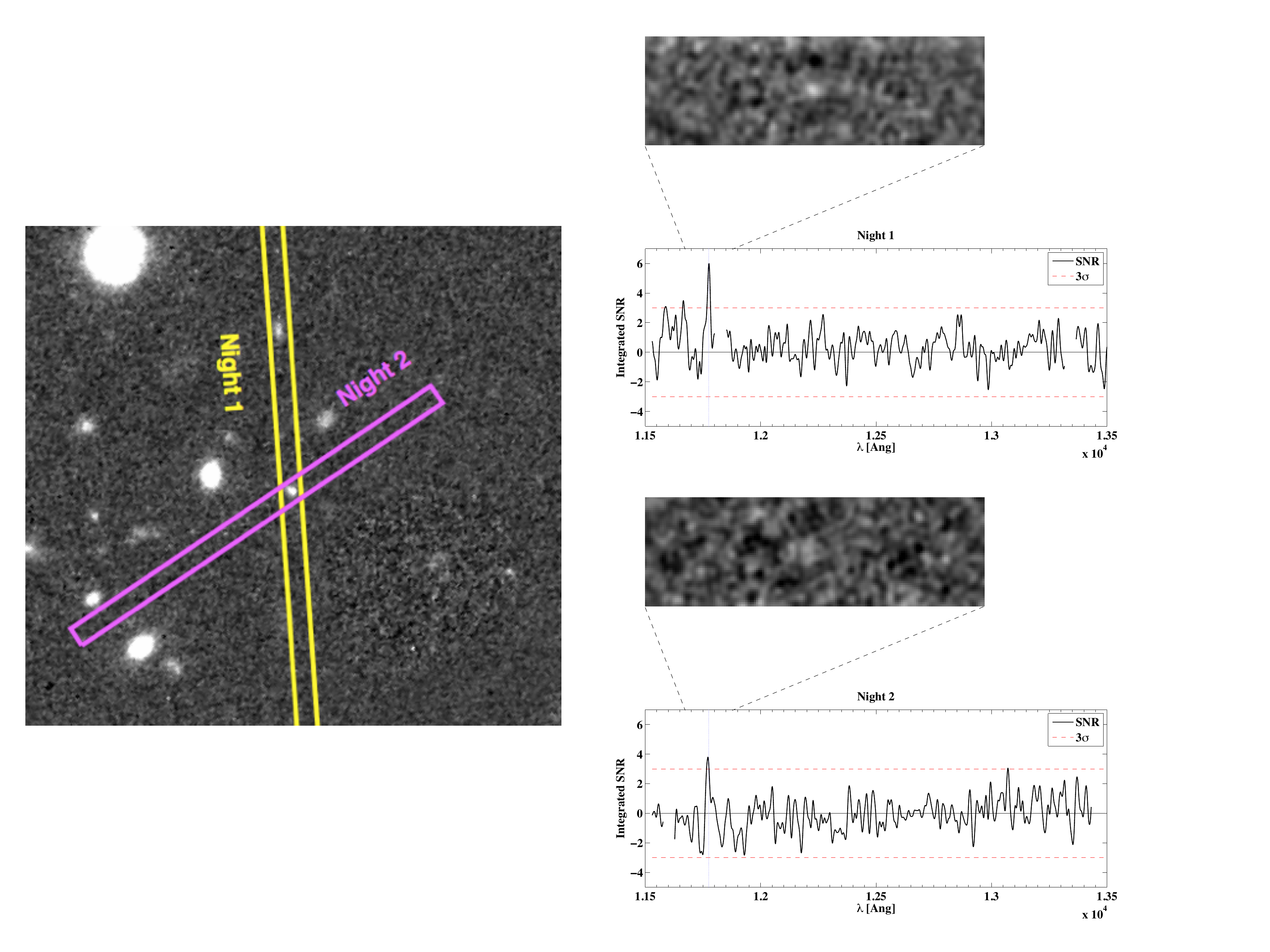}
 \end{center}
\caption{Confirmation of line detection in EGSY8p7 over two nights. \emph{Left:} \emph{HST} F160W image of the EGSY8p7 field with the slit orientations adopted in two successive nights. \emph{Right:} Extracted 1D signal/noise spectra across the J-band for each night, with a zoom of the 2D data around the line, marked with a vertical blue line. The 2D spectra are smoothed with a 3-pixel Gaussian for better illustration of the data. The 1D spectrum is smoothed with a Gaussian of $\sigma=5$ {\AA}, comparable to the measured line width. The Y axis is scaled so that the peak signal/noise matches the integrated value (e.g. \Finkelstein). Horizontal dashed lines mark the $\pm3\sigma$ region. \emph{On both nights, the signal/noise at the line location significantly exceeds that elsewhere}.}\vspace{0.1cm}
\label{fig2}
\end{figure*}

\section{Data} \label{obs}

The galaxy EGSY8p7 was detected in the Extended Groth Strip (EGS; \citealt{Davis2007EGS}) from deep ($\gtrsim27.0$) multi-band images in the CANDELS survey and first reported as one of four unusually bright ($H_{160}<25.5$) candidate $z>7$ galaxies by \RB. One of these, EGS-zs8-1, with $z_{phot}=7.92\pm 0.36$ was spectroscopically confirmed at $z=7.73$ by \Oesch. Normally such objects would be selected as $Y$-band dropouts but, given $Y$-band observations are not yet available over the full CANDELS field, an alternative selection criterion was adopted that takes advantage of red IRAC [3.6]-[4.5] colors, indicating prominent [O III]+H$\beta$ emission within the 4.5 $\mu$m band \cite[see also][]{Labbe2013,Smit2015EmissionLine}. Of the objects listed by \RB, EGSY8p7 is the only one for which Y-band data is not available. As a result, its photometric redshift derived from \emph{HST} data alone is fairly uncertain ($5.6<z_{phot}<9.2$; Fig. 5 of \RB) but including its IRAC [3.6]-[4.5] color of 0.76$\pm$0.14 (and blue $JH_{140}$-[3.6] color) narrows the range to $z_{phot}=8.57^{+0.22}_{-0.43}$.

We observed EGSY8p7, the highest-redshift candidate in \RB's list, on June 10 and 11 2015 with MOSFIRE on the Keck 1 telescope. Observations in the J-band spanned the wavelength range $11530 \AA < \lambda < 13520 \AA$ using an AB dithering pattern of $\pm1.25\arcsec$ along the slit with individual frames of 120s. The slit masks on the two nights differed by 120$^\circ$ in orientation both with a slit width of $0.7\arcsec$. In each mask we allocated one slit to a nearby star to monitor changes in seeing, transparency, and possible positional drifts. Conditions were clear throughout, with an average seeing of $0.60\arcsec$ for the first night and $0.76\arcsec$ for the second night. The total nightly exposure times were 158 minutes and 128 minutes. Excluding exposures where the seeing was significantly ($\gtrsim1\sigma$) worse than average, the useful exposure time comprises all frames from the first night and 80\% of those acquired on the second night (where the conditions were less stable) culminating in a total exposure-time of 4.33 hours. Calibrations were obtained via long-slit observations of standard A0V stars.

Data reduction was performed using the standard MOSFIRE reduction pipeline\footnote{http://www2.keck.hawaii.edu/inst/mosfire/drp.html}. For each flat-fielded slit we extracted the 1D spectrum using a 11-pixel boxcar centered on the expected position of the target. A similar procedure was adopted in quadrature to derive the 1$\sigma$ error distribution. To ensure that the derived error spectrum reflects the noise properties of the data, we measured the standard deviation of the pixel-by-pixel SNR, which should be unity for a set of independent values drawn from Gaussian distributions. We obtained a value of 1.3, implying that the errors are slightly underestimated by the pipeline, and corrected the error spectrum accordingly. Data from both nights were co-added by inverse-variance averaging the calibrated 1D spectra. We also allowed for relative shifts along the slit of 2 pixels ($\simeq$0.2\arcsec) and across the slit of 0.2\arcsec (which affects the expected slit loss correction) and propagated the associated uncertainty into our error budget. All reductions and calibration steps were performed independently by two authors (AZ, SB).

To calibrate the spectra we scaled a Vega model\footnote{http://kurucz.harvard.edu/stars/vega/} to each standard star to determine a wavelength-dependent \emph{relative} flux calibration and telluric correction using the procedure described in \citet{Vacca2003calibration}. Independent telluric calibrations derived from three standard stars agree to within $5\%$ at the location of the detected emission line. The \emph{absolute} calibration for each night was derived by comparing the spectroscopic magnitude of the stars on the slitmask with photometric measures in the $J$-band obtained from the 3D-HST catalog \citep{Brammer20123DHST,Skelton20143DHST}. Note that this procedure takes into account differential slit losses due to varying seeing. The absolute calibration factors obtained for the two nights are consistent with differences arising from seeing effects. We adopt a conservative absolute calibration error of $30\%$, taking into account the nightly variation, differences between the independent reductions, and other contributions mentioned above. The final 2D and 1D spectra are shown in Fig. \ref{fig1}.
\vspace{0.1cm}

\section{Lyman $\alpha$ at a Redshift 8.68}\label{lya}

Fig. \ref{fig1} reveals a prominent line at $\lambda\simeq11776$ {\AA} flanked by a skyline on its blue side. Examining the signal and associated noise in a circular aperture of 6 pixels in radius -- corresponding to $\sim2$ times the line width found below -- we estimate a significance of $7.6\sigma$ in the 2D spectrum and, within the extracted 1D spectrum over the same spectral range, $7.5\sigma$. This significance holds over a range of integration wavelengths both including and excluding the skyline. Additionally, the line is detected independently on each night ($6.0\sigma$ on night 1, $3.8\sigma$ on night 2), despite the changed slit orientation (Fig. \ref{fig2}). Taking the signal/noise on the first night, and assuming it scales as $\sqrt{t}$/FWHM, where $t$ is the exposure and FWHM the seeing, for the second night (where the exposure was less and seeing worse) we predict a signal/noise similar to that observed.

We estimate the observed line properties by using a Monte Carlo Markov Chain to fit a truncated Gaussian to the data (in a similar way to \Oesch), taking into account instrumental broadening. We leave the truncation point a free parameter, and mask out pixels contaminated by strong skyline residuals. The proximate skyline significantly affects the fit, and different masking configurations yield different results. To span the range of possible solutions we repeat the model fitting, as well as direct integration measurements, with a variety of masking configurations. In each such configuration a different part of the skyline was excluded. For each of these resulting models we calculate the 95\% confidence level intervals. For each parameter we then take the union of the intervals from different models and report them in Table 1. 

The line may be asymmetric with some attenuation on its blue side, but this is difficult to ascertain given the proximate skyline, and the best-fit models (Fig. \ref{fig1}) are mostly symmetric. The line peak lies between 11770 and 11776 {\AA} which, for Ly$\alpha$, implies a redshift of  $8.683^{+0.001}_{-0.004}$. Note the typical peak wavelength is a few {\AA} bluer than the raw peak in Fig. \ref{fig1}. The line width ranges from 2.0--8.4 {\AA}, corresponding to a velocity broadening, corrected for instrumental effects, of V$_{\sigma}$=51--214 km s$^{-1}$. The line flux ranges from 1.0--2.5$\times10^{-17}$ erg s$^{-1}$ cm$^{-2}$ (excluding the 30\% uncertainties in the absolute calibration). \emph{Importantly, the significance of the line remains at the $\gtrsim$7$\sigma$ level for all chosen integration methods and variations in the masking of the skyline}. The reliability of the SNR measurement given the nearby skyline was further verified using simulated lines inserted into the 2D spectra.

\begin{deluxetable}{lll}
\tablecaption{Emission Line properties}
\tablecolumns{2}
\tablewidth{0.85\linewidth}
\startdata
Name	& Fiducial model & 95\% C.I. \\
\hline
\\
f(Ly$\alpha$)\tablenotemark{a}		&  1.7 & [1.0 - 2.5]\\
$\mu$\tablenotemark{b}		&  11774 \AA  & [11770 - 11776] \AA \\
z\tablenotemark{c}	  &  8.683  & [8.679 - 8.684]  \\
{EW\tablenotemark{d}} &  28 \AA & [17 - 42] \AA  \\
$\sigma$\tablenotemark{e}  &   4.7 \AA &  [2.0 - 8.4] \AA \\
$V_{\sigma}$\tablenotemark{f}  &  118 km~s$^{-1}$ &   [51 -214] km~s$^{-1}$ \\
$V_{FWHM}$\tablenotemark{g}  &  277 km~s$^{-1}$ &   [120 -503] km~s$^{-1}$ \\
\enddata
\tablecomments{\\Emission line properties derived from a truncated Gaussian fit (Fig. \ref{fig1}), corrected for instrumental broadening, and not corrected for IGM absorption. The 95\% range quoted is adopted from a set of fits as described in the text.}
\tablenotetext{a}{Total flux in units $10^{-17}$ erg s$^{-1}$ cm$^{-2}$}
\tablenotetext{b}{Peak of Gaussian}
\tablenotetext{c}{Redshift}
\tablenotetext{d}{Rest-frame EW}
\tablenotetext{e}{Gaussian line width}
\tablenotetext{f}{Velocity width, corresponding to the Gaussian width}
\tablenotetext{g}{Velocity FWHM ($\simeq2.35\times\sigma$)}

\label{tab:specsummary}
\end{deluxetable}

\begin{figure}
   \includegraphics[width=90mm,clip]{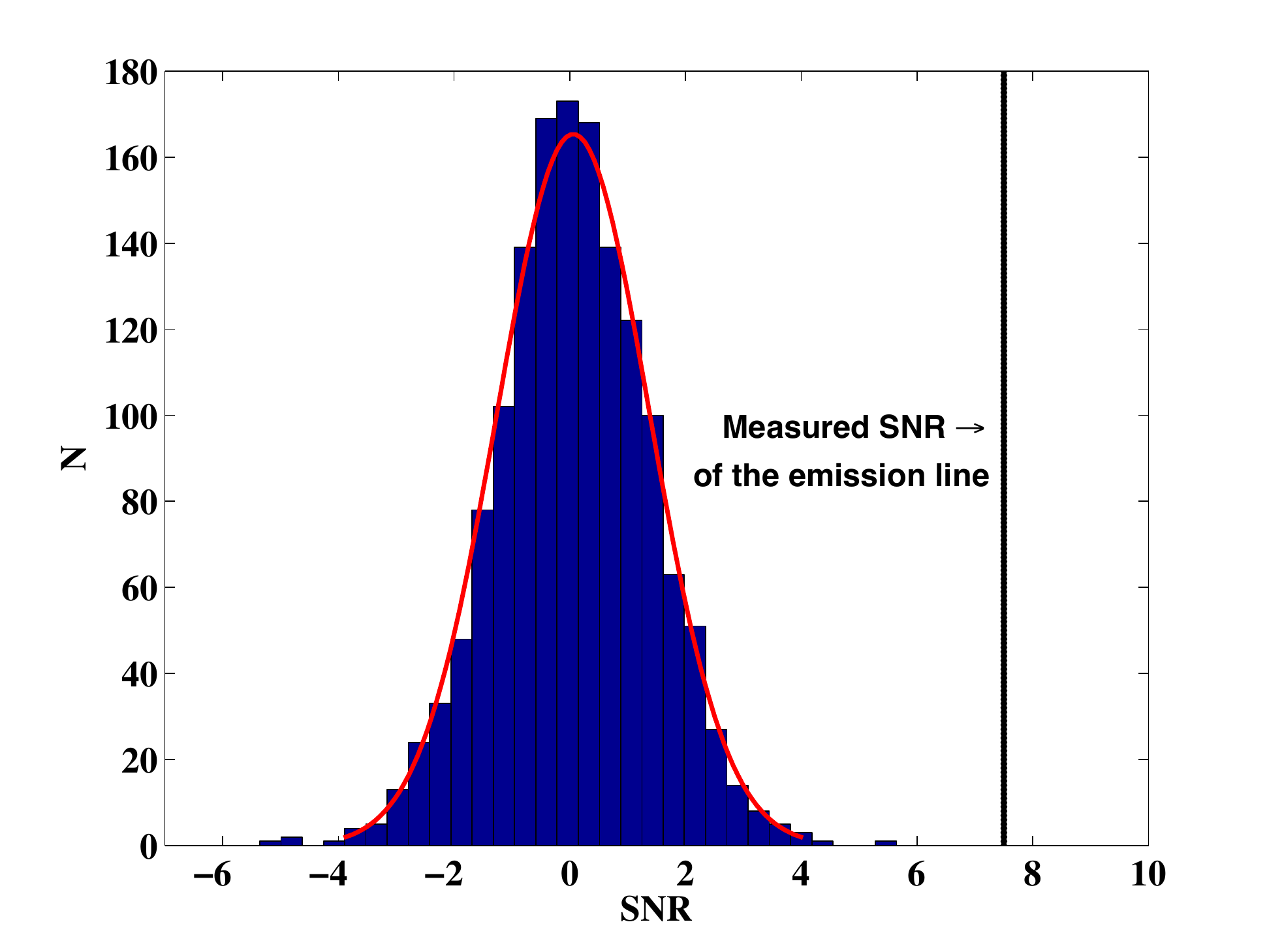}
\caption{Verifying the reality of the line detection in EGS8p7. The signal/noise distribution from 1500 randomly-positioned apertures across the 2D data presented in Fig. \ref{fig1}. All apertures were 6 pixels in radius ($\simeq7.5$ {\AA} in wavelength space), as the aperture used to measure the line significance. The vertical line shows the detected signal/noise far exceeds that of the noise distribution.}\vspace{0.05cm}
\label{random}\vspace{0.1cm}
\end{figure}

We can estimate the chance of finding such a line by examining the signal/noise distribution in our combined 2D spectral data at 1500 random locations taking the same aperture used for measuring the line properties above. The resulting histogram in Fig. \ref{random} reveals a very low probability of finding such a feature by chance.Our measured signal/noise represents a 6$\sigma$ deviation in the distribution (which has a width larger than unity due to correlated noise). We also verified that proximity to skylines does not increase the width of the distribution. Together with the facts that the positive and negative line positions in the 2D spectrum match the dithering pattern used (Fig.\ref{fig1}) and the line position is astrometrically centered on the target position along the slit on both nights, we conclude the probability for the detection to be an artifact is negligibly small.
 
Adopting the identification of Ly$\alpha$, the inferred redshift of  $z=8.683^{+0.001}_{-0.004}$ is satisfactorily close to the photometric estimate, $z_{phot}=8.57^{+0.22}_{-0.43}$ reported by \RB. Using the \emph{HST} $H$-band photometry unaffected by line emission, for a reasonable UV slope ($\beta=-2$), the range of line fluxes (Table 1) translates to a rest-frame EW of 17--42 {\AA}. While formally fairly uncertain, it is comparable to line properties secured with MOSFIRE for the other bright galaxies at $z_{phot}\gtrsim7.5$ selected in CANDELS with red IRAC colors (\Finkelstein, \Oesch). \Oesch\ obtained a total line flux of $1.7\pm0.3\times10^{-17}$ erg s$^{-1}$ cm$^{-2}$ and a rest-frame EW of $21\pm4$ {\AA}, whereas \Finkelstein\ secured a total line flux of $2.64\times10^{-18}$ erg s$^{-1}$ cm$^{-2}$ and a more modest rest-frame EW$\sim$7.5 {\AA}. We note \RB\ discussed the possibility that EGSY8p7 is magnified by foreground galaxies. Adopting the lens galaxy masses found in \RB\ as input, and using an updated photometric redshift, we conclude EGSY8p7 is likely magnified by $\sim20\%$, and less than a factor $\sim2$. 

\begin{figure}
   \includegraphics[width=90mm,trim=1cm 0cm 1cm 0cm,clip]{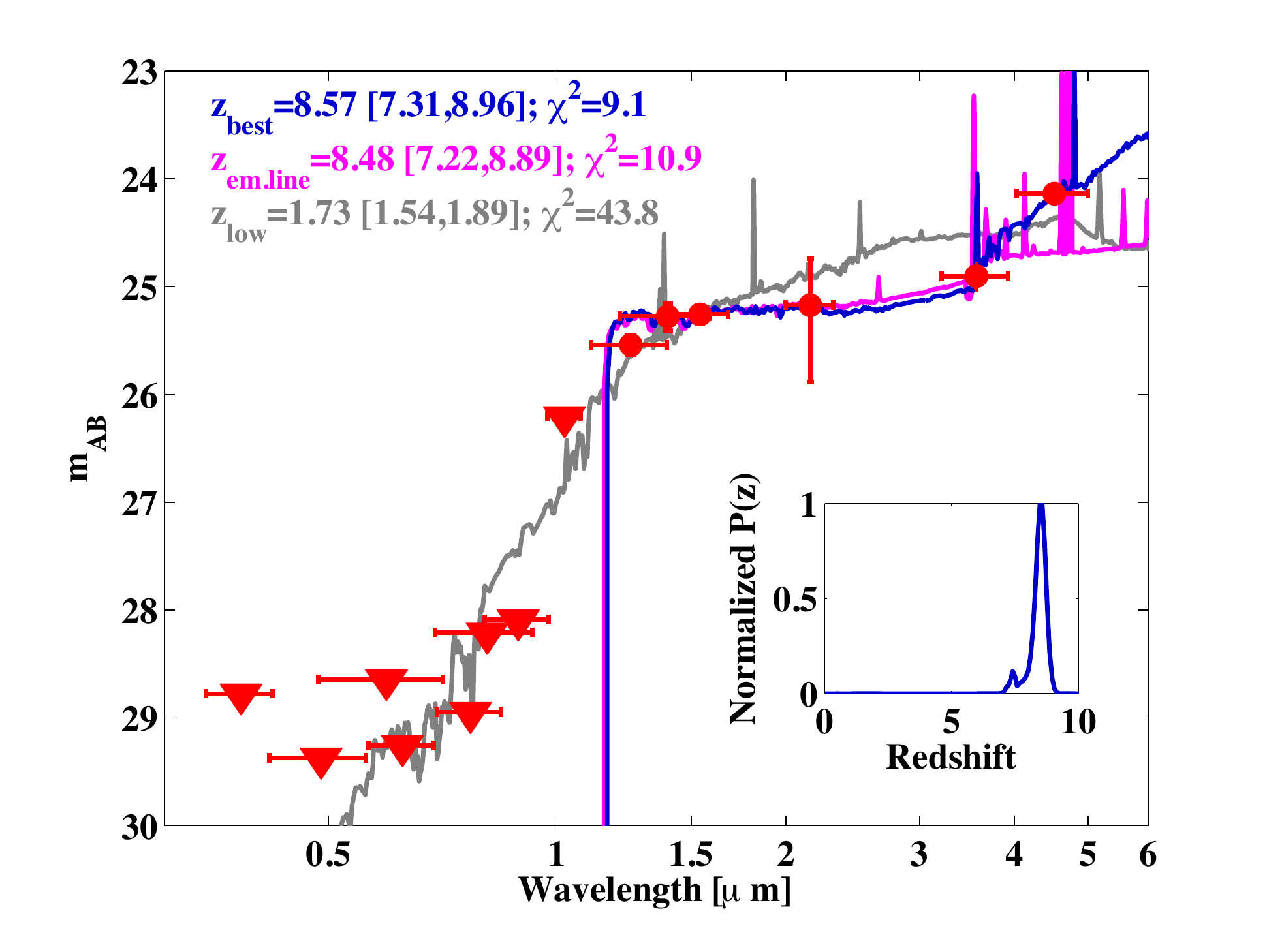}
\caption{Photometric SED fitting strongly supports a high-redshift solution for EGSY8p7: \emph{Red} points show magnitudes, where the horizontal error marks the filter width, and triangles denote 1$\sigma$ upper-limits from non-detections. The \emph{blue} SED shows the best-fit template at $z$=8.57. The \emph{magenta} SED shows the $z$=8.48 solution adopting the strong emission line template used in \RB\ (see their work for more details on the templates). The \emph{grey} SED shows the best-fit low redshift template which cannot explain the absence of detection in the optical bands and which is $\sim10^{7}$ times less likely than the high-redshift solutions.}
\label{fig4}
\end{figure}

Given we only detected one spectral line in the J-band, we also considered interpretations other than Ly$\alpha$, corresponding to a lower redshift galaxy. If the line is a component of the [O II] ($\lambda\lambda$3726, 3729) doublet, this would imply $z\simeq2.16$, possibly consistent with a low redshift solution to the SED (Fig. \ref{fig4}; $z_{phot}=1.7\pm0.3$; 99\% C.I.). However, for each [O II] component we find no trace of the other line to a limit several times fainter than the detected line (\emph{magenta} arrows in Fig. \ref{fig1}). Similarly, in the case of [O III]  ($\lambda\lambda$4959, 5007; \emph{cyan} arrows), or the H$_{\beta}$ line ($\lambda4861$), implying $z\sim1.4$, we would expect to see (at least) one of the other lines given typical line ratios assumed. While the presence of skylines clearly limits a robust rejection of these alternatives, both the absence of any optical detections and the red IRAC color are hard to understand in a low redshift galaxy with intense line emission. Formally, the SED fit, performed using the EAZY program \citep{Brammer2008EAZY}, gives a low redshift likelihood $10^{7}$ times smaller than our adopted solution at $z=8.68$ (Fig. \ref{fig4}). 


\section{Discussion and Summary} \label{discussion}

Although we claim EGSY8p7 is the highest-redshift spectroscopically-confirmed galaxy and the first star-forming galaxy secured beyond a redshift of 8 (c.f. a gamma ray burst at $z=8.2$, \citealt{Tanvir2009}), its prominent Ly$\alpha$ emission, detected in only 4 hours, raises several interesting questions about the usefulness of galaxies as tracers of reionization. The last few years has seen a consistent picture emerging from observational programs charting the visibility of Ly$\alpha$ beyond a redshift $z\simeq$6.5 in much fainter ($m\simeq27$) galaxies (\citealt{Schenker2014above7Spec} and references therein). Although the exposure times, selection techniques and observational strategies differ across the various programs, almost 100 faint targets have now been targeted, collectively yielding very few $z>7$ Ly$\alpha$ detections (\Finkelstein, \citealt{Schenker2014above7Spec}). In marked contrast, all 3 bright $z_{phot}\gtrsim7.5$ targets in the EGS field in Table 2 of \RB's compilation now reveal Ly$\alpha$ emission.

Although this bright spectroscopic sample is admittedly modest, the high success rate in finding Ly$\alpha$ at redshifts where the IGM is expected to be substantially neutral ($\simeq$70\% according to \citealt{Robertson2015}) raises two interesting questions. Firstly, do these new results challenge claims made by the earlier fainter Ly$\alpha$ searches in the same redshift interval? Is it possible that those campaigns reached to insufficient depths to reliably detect Ly$\alpha$ emission? Typically such studies adopted the rest-frame EW distribution at $z\simeq$6 published by \citet{Stark2011LAE} and simulated, using their appropriate survey and instrumental parameters, the likely success rate to somewhat higher EW limits of $\simeq$25-50 {\AA}. A key issue in understanding whether the \RB\ objects are similarly attenuated by the partially neutral IGM at $z\gtrsim7.5$ is the fiducial $z\simeq$6 EW distribution for these luminosities. Stark et al's EW distribution is based on an analysis of 74 $z\simeq6$ galaxies targeted spectroscopically within the deep GOODS fields but only 13 reached the UV luminosities being probed here. \citet{Curtis-Lake2012UDS} present a smaller sample of 10 spectroscopically-confirmed $z\simeq$6 galaxies from the UDS field, several of which are more luminous with EWs $\gtrsim$ 40 {\AA}. A key deficiency in this comparison is that the $z\simeq6$ samples were not selected to have as large EW [OIII]+H$\beta$ emission implied in the IRAC [3.6]-[4.5] color-selected \RB\ sample. \citet{Stark2014z2CIIILymanalphaZ2} show that Ly$\alpha$ EWs are uniformly extremely large (40 --160 \AA) in $z\simeq$2-3 galaxies with such large EW optical line emission. Thus it is possible that Ly$\alpha$ in these $z_{phot}\gtrsim7.5$ galaxies is attenuated. In addition to the possibility of cosmic variance in the EGS field, it may be that such luminous objects at $z>7$ lie in large overdense regions and present a somewhat accelerated view of the reionization process \citep{BarkanaLoeb2006}.

The second question is whether these rarer luminous sources (c.f. \citealt{Matthee2015BrightLAEz67,Ouchi2013HimikoZ6p6}) are physically distinct from the fainter population. Adopting the IRAC red [3.6]-[4.5] color selection may preferentially select unusually intense line emission with abnormal ionizing spectra capable of creating early bubbles of ionized gas in the local IGM (see also \Finkelstein, \citealt{Ono2012}). Evidence for stronger than usual ionizing spectra in $z>7$ dropouts with known Ly$\alpha$ emission is provided by the prominent detection of CIV 1548 \AA\ emission in the lensed galaxy at $z=7.047$ \citep{Stark2015CIV}. The key issue in this case is the proportion of $z>6.5$ Lyman break galaxies that reveal such prominent IRAC excesses. \citet{Smit2015EmissionLine} demonstrate that perhaps upwards of 50\% of the photometrically selected galaxies at $z\simeq$6.6-6.9 have IRAC excesses that are comparable in magnitude to those seen in \RB. 

In summary, not only have we pushed the detectability of Ly$\alpha$ emission to well beyond a redshift $z\simeq$8, but it seems luminous $z_{phot}>7.5$ galaxies selected additionally for their red IRAC [3.6]-[4.5] colors have an unusually high success rate of line detection. We present several reasons why this might contrast with the low yield of detecting Ly$\alpha$ in intrinsically fainter sources. Regardless of the explanation, it is clear that further spectroscopic follow-up of such examples will yield valuable information on the early development of massive galaxies.

\vspace{0.12cm}

\section*{acknowledgements}
We thank the reviewer of this work for useful comments. We acknowledge useful discussions with Jim Dunlop, Brant Robertson and Chuck Steidel. We thank Garth Illingworth, Pieter van Dokkum, Iva Momcheva, Marijn Franx, Mauro Stefanon, and Benne Holwerda for their role in the selection of this object. Support for this work was provided by NASA through Hubble Fellowship grant \#HST-HF2-51334.001-A awarded by STScI, which is operated by the Association of Universities for Research in Astronomy, Inc. under NASA contract NAS~5-26555. RSE acknowledges the hospitality of the Institute for Astronomy, Edinburgh via a Carnegie Centennial Professorship. This work is in part based on previous observations made with the NASA/ESA Hubble Space Telescope, the Spitzer space telescope, the Subaru telescope, and the Canada France Hawaii Telescope. The data presented herein were obtained at the W.M. Keck Observatory. The authors wish to recognize and acknowledge the very significant cultural role and reverence that the summit of Mauna Kea has always had within the indigenous Hawaiian community. We are most fortunate to have the opportunity to conduct observations from this mountain.

\vspace{0.12cm}


\end{document}